# Deterministic multi-level spin orbit torque switching using focused He$^+$ ion beam irradiation

*Jinu Kurian[1], Aleena Joseph[1], Salia Cherifi-Hertel[1], Ciaran Fowley[2], Gregor Hlawacek[2], Peter Dunne[1],*

*Michelangelo Romeo[1], Gwenaël Atcheson[3], J. M. D. Coey[3], Bernard Doudin[1*]*

[1]Université de Strasbourg, CNRS, IPCMS UMR 7504, 23 rue du Loess, F-67034 Strasbourg, France

[2]Institute of Ion Beam Physics and Materials Research, Helmholtz-Zentrum Dresden - Rossendorf, Bautzner Landstraße 400, 01328 Dresden, Germany

[3]AMBER and School of Physics, Trinity College, Dublin 2, Ireland

* Corresponding author; email: bernard.doudin@ipcms.unistra.fr

He$^+$ ion irradiation is used to pattern multiple areas of Pt/Co/W films with different irradiation doses in Hall bars. The resulting perpendicular magnetic anisotropy landscape enables selective multilevel current-induced switching, with full deterministic control of the position and order of the individual switching elements. Key pattern design parameters are specified, opening a way to scalable multilevel switching devices.

Spin orbit torque (SOT) magnetic memory devices have unique advantages in terms of low power consumption, fast operation capabilities and simplified circuitry design, placing them at the forefront of information technology development.[1–3] Their current-induced switching properties rely on the perpendicular magnetic anisotropy (PMA) of their thin magnetic layers, which is readily modified by light ion irradiation.[4] We have previously shown the benefit of combining nanometer-scale irradiation in a He$^+$ microscope with *in situ* electronic transport property measurements to reveal how the local PMA evolves under irradiation.[5] Reduction of the PMA in specific locations makes selective SOT switching of magnetic bits possible. Our purpose here is to show how this technique can be extended to multiple irradiation zones in the same device, taking advantage of fine tuning the spatial distribution of PMA calibration, to achieve multi-level switching in devices.

Multi-level storage is a promising approach to increase memory density[6] while avoiding the addressability problems associated with size reduction. Increasing the number of available states by a factor *N* leads to an *N*-fold improvement in the density of memory without any change to the device geometry. Multi-level SOT devices can also be exploited for bio-inspired computing architectures, such as neuromorphic computing based on spintronics[7,8], and multi-level switching has been achieved in PMA heterostructures in many ways. These include stabilizing intermediate multidomain states by means of multi-ferromagnetic layer stacks,[9–12] wedged magnetic layers,[13,14] ferrimagnetic[15,16], ferromagnetic/antiferromagnetic structures[17], or with interleaved interfaces with different spin Hall angles.[18] A different approach defines the multiple levels as a sequence of pinning-depinning states in SOT-driven reversal[19,20], which has been shown in both irradiated and non-irradiated single magnetic films[21]. Here, several unique states are accessible using shaped current pulses[22], but the exact magnetic configuration is highly dependent on sample fabrication and reproducibility.[23] Our aim is to create well-defined, reproducible, deterministic states, that are addressable by SOT switching, and avoid the limitations of sample reproducibility.

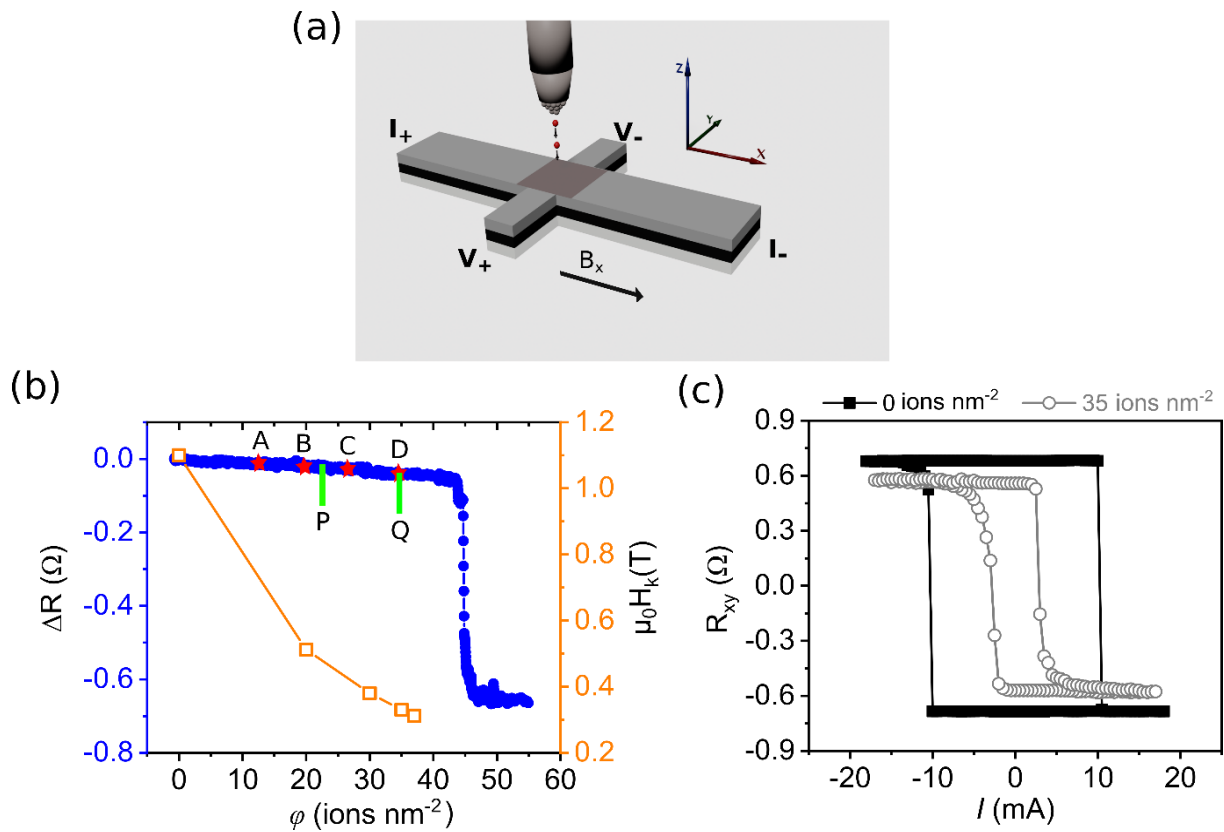

Figure 1. (a) Schematic of SOT setup of a patterned Hall bar undergoing He$^+$ irradiation with the directions of current and magnetic field indicated. (b) Evolution of *in-situ* anomalous Hall resistance with irradiation, the letters P,Q indicate the two-zone irradiation doses shown in Fig. 2a, and A-D the four-zone doses shown in Fig. 3a. The magnetic anisotropy field (orange squares) as a function of irradiation, deduced from Hall measurement on *ex-situ* samples, is also indicated. (d) Comparison of SOT induced magnetization switching before and after irradiating with 35 ions/nm$^2$ under a 100 mT bias field.

Thin film stacks of Ta (5.0)/Pt (2.0)/Co (1.2)/W (1.5)/Pt (1.5)/Ta (1.5) (thickness in nm) were grown on thermally-oxidized Si substrates by DC magnetron sputtering at room temperature. The Ta (5.0) seed layer ensure good adhesion and (111) texture of Pt that promotes perpendicular anisotropy of the Co layer.[24] The heterostructure of Co sandwiched between Pt and W was chosen to optimize SOT switching efficiency due to opposite signs of the spin Hall angles of Pt and W. The PMA of the cobalt was confirmed by magnetometry. Films were patterned into Hall bar structures of width 10 μm and length 180 μm by UV lithography. The Hall crosses were irradiated using an Orion Nanofab helium ion microscope system while monitoring the evolution of the anomalous Hall resistance ($R_{AHE}$) *in-situ*. (Figure 1 a,b)[5]. For uniformly irradiated Hall crosses, $R_{AHE}$ gradually decreases with irradiation until there is a sharp drop at a critical dose of 42.5 ± 2 ions/nm$^2$, where the easy axis of magnetization flips from out-of-plane to in-plane. The critical dose is reproducible within 5% from sample-to-sample.

The magnetic anisotropy field (Fig. 1b) was measured *ex-situ* by Hall effect as a function of applied field. Hall measurements were performed under 1mA bias current. SOT-induced switching experiments were performed with 1 ms current pulses in an interval of 1 s, under 100 mT bias field applied along the current direction (Fig. 1c) . An average current density values of approximately 8 MA/cm$^2$ corresponds to a 10 mA pulse current. Our experimental procedure does not give indications of long-term sample Joule heating if the current stress remains below typically 35 mA. One can imagine Joule heating during the 1 ms current pulse, but we found that the current hysteresis curves remain fully reproducible (within a few percent) if the current sweeps does not exceed 35 mA, providing confidence that the current stress does not impact the Co thin film and its interface down the atomic level. The observed trends in the critical dose, fall in anisotropy and reduced critical current are consistent with our previous work.[5] Fig. 1c compares the SOT switching hysteresis curve of a non-irradiated cross and one irradiated, illustrating the reduction of switching current after irradiation with 35 ions nm$^{-2}$ Further characterisation and calibration of both the magnetic anisotropy as well as SOT switching currents was carried out for several doses below the critical dose. In this work, we focus on Hall crosses which were individual well-defined zones were irradiated with different doses.

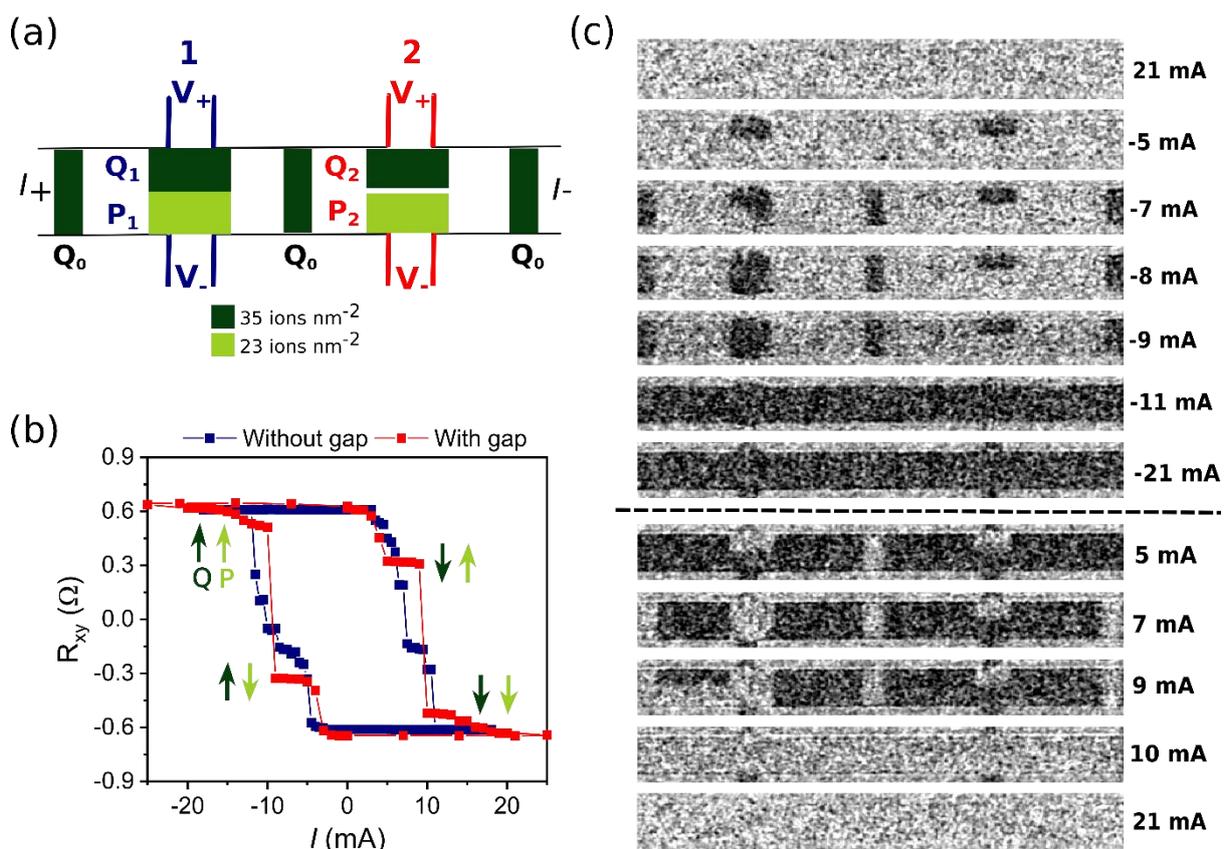

Figure 2. (a) Schematic of a two-zone irradiated sample (doses P,Q Fig. 1b) with and without a gap between irradiated zones, and two boundary zones Q$_0$ to minimize domain wall propagation during SOT switching (see text). (b) SOT induced magnetization switching of two junctions with and without a gap between two irradiated zones with dose 23 ions/nm$^2$ and 35 ions/nm$^2$ at a bias field of 125 mT. (c) MOKE images of the SOT induced magnetization switching when sweeping the current from +21 mA to -21 mA, and then (below broken line) back to +21 mA. Intermediate currents (21 to -5 mA, and -21 to 5 mA) where the magnetization does not switch are not shown.

Fig. 2 details the simplest case of a two-zone irradiation device, with one zone irradiated with 23 ions nm$^{-2}$ (dose denoted 'P' in Fig. 1b) and the other with 35 ions nm$^{-2}$ ('Q'). For a direct comparison we irradiated neighbouring Hall crosses, labelled 1 and 2 in Fig. 2a. In cross 1 the irradiated zones are adjacent to each other, whereas in cross 2 they are separated by approximately 1 µm. In both crosses, we detect additional resistance states due to the different magnetic anisotropies of the two zones P & Q, denoted as ↑↑, ↑↓, ↓↑ and ↓↓ (Fig. 2b). These states have unique $R_{AHE}$ values and are accessible by cycling the applied current pulses. We found that measured Hall resistance values were stable within 1% over many hours. The resulting states are non-volatile, holding their resistance values for at least two days. The switching currents of the irradiated areas are also reproducible within a few percent when repeating the current sweeps. Magneto-optical Kerr effect (MOKE) microscopy confirms the correspondence between the Hall resistance data and the switching of each irradiated area, allowing us to visualize the resulting magnetic configurations (Fig. 2). The boundary of non-irradiated and irradiated zones are known to pin propagating domain walls during magnetic field driven reversal[25,26], and are found to play such a role in the SOT induced switching by Kerr microscopy visualisation (not shown here). We found that it was crucial to irradiate magnetic boundary zones on either side of the Hall cross, marked '$Q_0$' in Fig. 2a, using a near-critical dose so that these zones switch at the smallest currents. This inhibits propagation of magnetization reversal from far away in the main bar into the cross, which would interfere with the Hall measurement. Essentially, this additional irradiation acts to isolate the magnetic switching behaviour of a cross, without severing its electrical contacts. Unfortunately, we found that reproducibility of the switching current of unirradiated zones surrounded by $Q_0$ irradiated areas was quite poor, but nevertheless sufficient to avoid a nucleation of reversal far from the sample of interest at lower current values than those switching irradiated areas.

Fig. 2 also illustrates how independent switching of the two zones P and Q can be improved by leaving a non-irradiated strip between them. A width of 1 µm was chosen to allow adequate resolution in the Kerr microscope. Under a bias field of 125 mT we find a clear independent SOT switching of P and Q in both the measured Hall resistance (Fig. 2b) and MOKE imaging (Fig. 2c) only when there is an irradiation gap ($P_2$, $Q_2$). When the current is swept from 21 to -21mA, the zones irradiated with 35 ions nm$^{-2}$, $Q_1,Q_2$, selectively switch at -5 mA and the adjacent irradiated zone, $P_1$, reverses its magnetization continuously between -7 mA and -8 mA. In contrast, the zone $P_2$, separated by the non-irradiated gap, switches at -11 mA. This larger critical current results from domain wall pinning in the unirradiated zone which retains the larger initial PMA, preventing the reversal of one zone to expand to the neighbouring one. Note that for this sample, the unirradiated areas switch at current values similar to $Q_2$, which explains why the antiparallel state does not correspond to zero $R_{AHE}$.

Irradiation of multiple zones within a Hall cross illustrate the scalability of this process. Four zones, A–D, separated by 1 µm gaps and irradiated with different doses, 35, 27, 20 and 13 ions nm$^{-2}$, are shown in Fig. 3a. As with the sample shown in Fig. 2, magnetic isolation bands, $Q_0$, were irradiated either size of the junction with a dose of 35 ions nm$^{-2}$, protecting the unirradiated zone against switching up to 20 mA current. The SOT induced switching of the device under a bias field of 125 mT in Fig. 3b shows the three intermediate steps corresponding to the independent switching of the four zones upon sweeping applied current. MOKE microscopy confirms this interpretation of the AHE data (Fig. 3c), displaying a sequential switch of each area from lowest to highest PMA, i.e. from highest to lowest ion dose. This demonstrates that our approach can be extended to irradiate *N* zones with different doses, across different areas. We expect the domain wall width to define the scale of the minimum bit size, leading to a separation wall of a few tens of nm for individual cells below the 100 nm size range. This is well within the resolution limits of the irradiation process but is challenging for magnetic imaging and electrical detection and is beyond the scope of this paper.

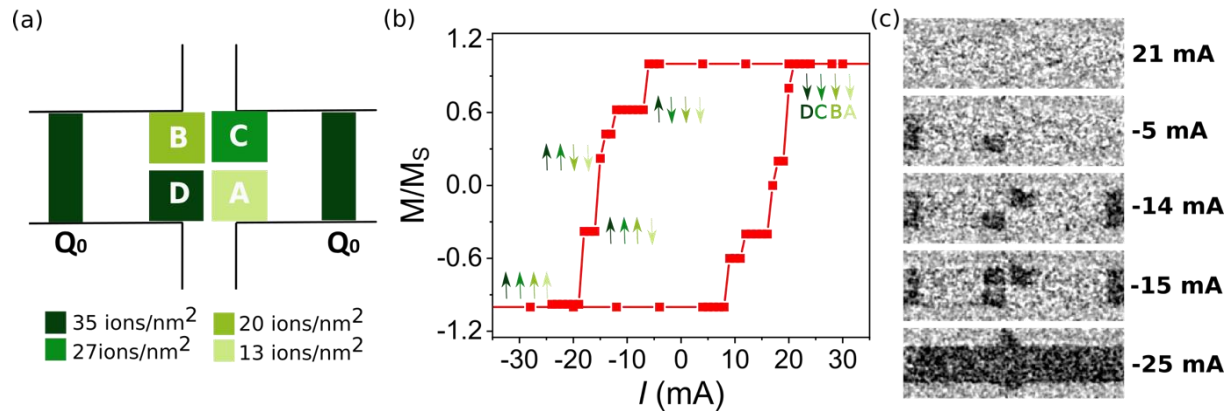

Figure 3. (a) Schematic of the four zones of a junction irradiated with doses 13 ions nm$^{-2}$, 20 ions nm$^{-2}$, 27 ions nm$^{-2}$ and 35 ions nm$^{-2}$. (b) MOKE acquired SOT induced magnetization switching of the junction under a bias field of 125 mT. (c) MOKE images of the SOT induced magnetization switching of the junction when sweeping from 21 to – 25 mA.

In summary, magnetic anisotropy engineering by appropriate multi-zone He-ion irradiation makes multi-level switching of magnetic configurations robust and predictable. Our experimental design was chosen for maximum simplicity, to demonstrate that we can independently switch the magnetic state of individually irradiated areas in a single Hall cross. Moreover, by taking advantage of the hysteresis properties of the current-induced switching, many macrospin configurations become accessible (green arrows in Fig. 2b), and one can imagine switching any combination of the macrospin presented in Fig. 2. This would allow the number of distinct electrical states to be much larger than the number of irradiated zones $N$. More sophisticated designs are also possible on a smaller scale, for example where a switching of a given area influences its neighbour via its magnetic stray field. Furthermore, ion irradiation is compatible with other approaches for realizing multistate memories without altering their physical design, such as more sophisticated magnetic stacks, wedge layers or lithographic patterns.[8-16] We believe that the strategy we have outlined proposes a versatile strategy to realize high-density and low-power-consumption SOT-based memory devices and is relevant for future types of computing architecture.

**Author Contributions**

P.D., C.F. and J.K. initially designed the experiment, J.K. and A.J. performed most experiments, G.A. and J.M.D.C. helped for the materials fabrication, G.H. C.F. for the ion irradiation, S.C. and M.R. for Kerr measurements. B.D. and J.M.D.C. supervised the experiments. J.K., A.J. and B.D. wrote the manuscript, where all authors contributed to its improvement.

**Acknowledgements**

We thank Fabien Chevrier and the staff of the STnano nanofabrication facility for daily support. This project has received funding from the European Union's Horizon 2020 research and innovation programme under the Marie Skłodowska-Curie grant agreement MaMi No. 766007 and QUSTEC No. 847471, the Interdisciplinary Thematic Institute QMat, as part of the ITI 2021 2028 program of the University of Strasbourg, CNRS and Inserm, was supported by IdEx Unistra (ANR 10 IDEX 0002), and by SFRI STRAT'US project (ANR 20 SFRI 0012) and EUR QMAT ANR-17-EURE-0024 under the framework of the French Investments for the Future Program.

**Conflicts of interest**

There are no conflicts of interest